# Miniature narrow-linewidth 1 μm Laser

Xiaofan Zhang[1], Fan Zhang[1], Kunpeng Jia[1], Yunfeng Liu[2], Haosen shi[3], Yanyi Jiang[3,4], Xiaoshun Jiang[1], Longsheng Ma[3,4], Wei Liang[2], Zhenda Xie[1], Shi-ning Zhu[1]

[1] *National Laboratory of Solid State Microstructures, School of Electronic Science and Engineering, College of Engineering and Applied Sciences, School of Physics, and Collaborative Innovation Center of Advanced Microstructures, Nanjing University, Nanjing 210093, China.*

[2] *Suzhou Institute of Nano-tech and Nano-bionics, Chinese Academy of Sciences, Suzhou 215123, China.*

[3] *State Key Laboratory of Precision Spectroscopy, East China Normal University, Shanghai 200062, China.*

[4] *Collaborative Innovation Center of Extreme Optics, Shanxi University, Taiyuan 030006, China*

Authors to whom correspondence should be addressed: [Kunpeng Jia, jiakunpeng@nju.edu.cn; Wei Liang, wliang2019@sinano.ac.cn; Zhenda Xie, xiezhenda@nju.edu.cn ]

Self-injection locking scheme has the potential to narrow the linewidth of lasers in a compact setup. Here, we report a narrow linewidth laser source near 1 μm by self-injection locking scheme using a Fabry-Perot (FP) hollow resonator with a high-quality factor ($Q>10^8$). The measured fundamental linewidth of the laser is 41 Hz, and a coarse tuning range over 5.5 nm is achieved by changing the driving current of the laser source. Meanwhile, a fine-tuning range of 373 MHz is achieved without mode hops by changing the voltage applied to the PZT on the resonator. More importantly, benefiting from the low thermal refractive noise and low thermal expansion of the FP hollow resonator, the beat-note linewidth and the frequency Allan deviation are measured to be 510.3 Hz in and $10^{-11}$ (1s averaging time), respectively, by using a fully stabilized frequency comb as reference. Such a high-performance laser is fully integrated with a palm-sized package (52.3 mL) for field-deployable applications.

Narrow linewidth lasers near 1 μm are essential for massive applications such as nonlinear frequency conversion, medical diagnosis, high-throughput laser material processing, interstellar laser interferometer (ILEI), etc. [1-5]. Self-injection locking scheme is well-known for its capability to considerably narrow down the linewidth of a diode laser. Benefiting from the simple geometry, the self-injection locking scheme has the potential to enhance the laser performance in a compact setup for practical applications [6-8]. Recent studies on self-injection locking lasers ranging from bulk Fabry-Perot (FP) resonators to on-chip micro-resonators have realized sub-Hz to kHz narrow fundamental linewidth [9-17]. Among these, the hollow FP resonator, with low thermal refractive noise and low thermal expansion, plays an important role in optical paths to achieve stable feedback and enhance noise performance, which is essential for applications such as atomic, molecular and optical (AMO) physics. To date, most of the self-injection locking research focuses on the optical communication wavelength band, due to the developed high-performance laser diodes and other optics working at this band. Narrow-linewidth 1 μm self-injection locking laser has rarely been demonstrated.

In this work, we demonstrate an ultra-narrow linewidth 1μm laser by self-injection locking of an electrically pumped laser chip to a high-quality (Q) factor ($>10^8$) FP hollow resonator. An ultralow-noise performance is measured with the white frequency noise of 13 $Hz^2$/Hz, corresponding to the fundamental linewidth of 41 Hz. By changing the driving current of the laser source, a course tuning range over 5.5 nm is achieved for the ultra-narrow linewidth laser near 1 μm output, and more importantly, a fine-tuning range of 373 MHz can be performed without mode hops. We characterize the beat frequency of our laser by beat-note frequency measurement using a fully stabilized frequency comb, which has a linewidth of 510.3 Hz and $10^{-11}$ frequency instability at 1 s averaging time. All the laser chips, the FP hollow resonator, and micro-optics are packaged in a miniature setup with a volume of 52.3 mL, which may facilitate a wide range of field-deployable applications [18]. This is a 1 μm self-injection locking laser with an integrated linewidth as low as 500 Hz and a palm-sized package.

In our sub-mL high-Q FP resonator, a commercial plane mirror and a concave mirror (radius of curvature of 250 mm) are the main elements that make the beam oscillate as shown in Fig. 1 (a). The center of the two mirrors is a 9 mm long hollow fused silica spacer, and the reflectivity of the mirrors is 99.6%. The transmission medium is air which features low thermal refractive noise and low nonlinear noise, keeping the resonator quiet. Although the FP resonator has many transverse modes, by carefully designing the input beam to match that of the fundamental mode, one can achieve nearly single-mode excitation. We characterized the FP hollow resonator by scanning the frequency of a 1 μm tunable semiconductor laser (CTL 1064, Toptica), a narrow resonance linewidth of 2.1 MHz is measured, finesse of 7952 and a Q factor of $1.34 \times 10^8$. Such a high-quality mode of an external resonator provides fast frequency-selective optical feedback, which leads to improved stabilization of the laser frequency and keeps the single-mode (SM) lasing [19-20].

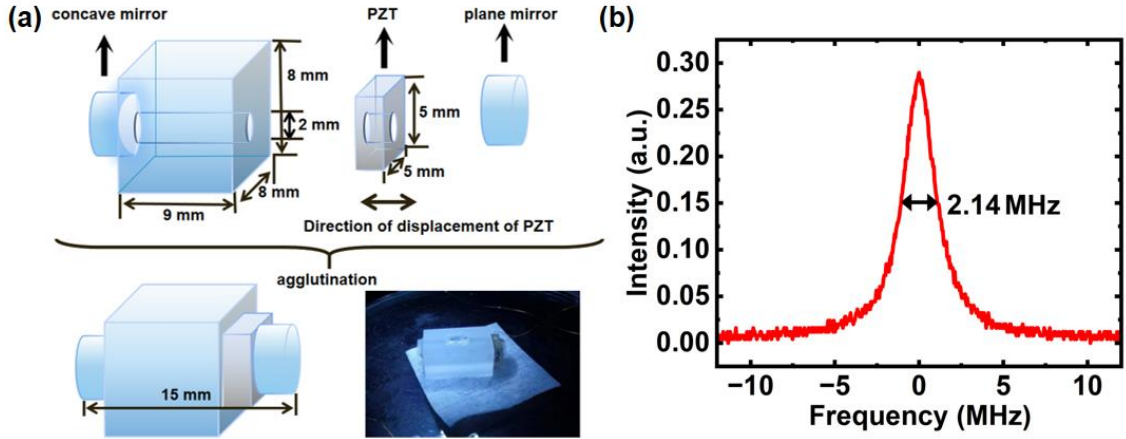

Fig. 1 (a) Picture of the FP resonator. (b) Characterization of the FP resonator. Measured transmission spectrum of the FP resonator using a tunable laser.

The experimental setup of the self-injection locking laser is schematically shown in Fig. 2 (a). When the laser source (Eagleyard, EYP-RWS-1064-00080-1500-ALN26-0000) frequency is aligned sufficiently close to a resonance frequency, the optical field reflected from within the resonator provides feedback to spontaneously align the laser to the resonance, dramatically reducing the laser linewidth [21-24]. The emission of the laser chip is collimated by the lens and passes through ports 1, and 2 of a free space circulator. It is then launched to an isolator (ISO) for better isolating back scattering light from the reflector of the FP resonator. After passing through the isolator, the reflector, and the half wavelength plate (HWP) successively, the collimated light is split into two parts by a polarizing beam splitter (PBS). The reflected part of the light is coupled into the FP resonator with its transmission fed back to the laser chip through port 3 of the free space circulator. The transmitted part of the light is coupled to polarization-maintaining fiber as output. In our self-injection locking scheme, the stability of our laser mainly depends on the length stability of the FP resonator. The HWP is used for polarization control before PBS. The laser chip is driven by an ordinary constant current source (Thorlabs, LDC200C) with an output current range of 50-100 mA. The total volume of the optical bench containing all optical components is only 52.3 mL after packaging. We use two temperature control systems (Thorlabs, TED200C) to stabilize the temperatures of the laser chip and the optic bench respectively.

We characterize the frequency tunability of the self-injection locking laser. When the laser frequency is tuned into the resonance of FP resonator, the diode laser is injection-locked with higher spectral purity and the maximum output power is 8 mW. Our laser can be tuned to 5.5 nm by changing the current, as shown in Fig. 2 (b). Such tuning capability is important for the tunable pump of OPO. Meanwhile, we added piezoelectric ceramics (PZT, PA4FKH3W, Thorlabs, Inc.) to the FP hollow resonator to realize fine-tuning of the laser frequency. By squeezing the resonant cavity, the PZT affects the length of the cavity length, which makes the resonance conditions change. An arbitrary function generator is used to generate an external drive triangular wave electrical signal applied to PZT to change resonance conditions. The continuous frequency tuning range of the self-injection locking laser is about 373 MHz with the PZT voltage range of 930 mV, as shown in Fig. 2 (c).

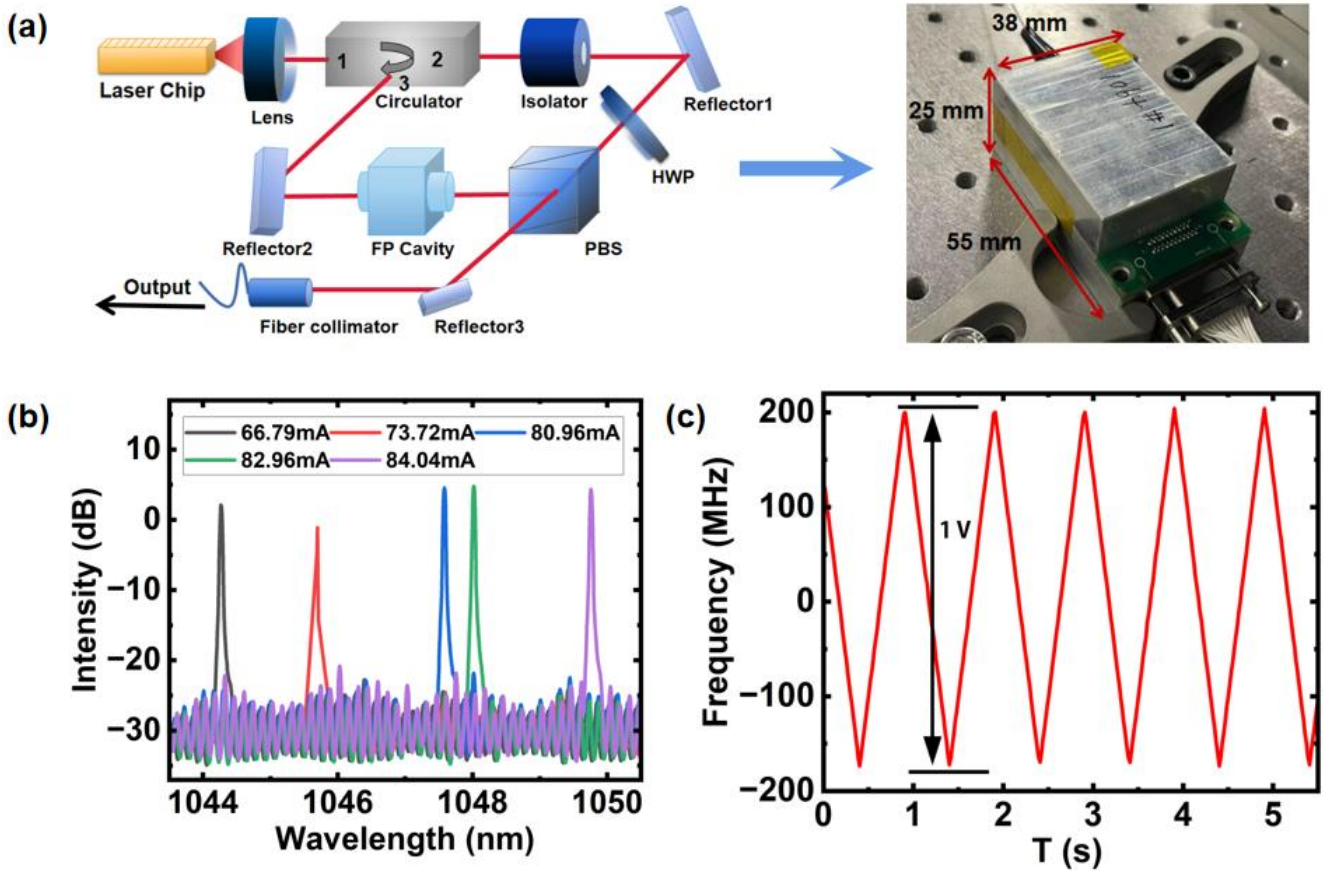

Fig. 2 (a) Setup of the self-injection lock laser. (b) Optical spectra showing the tunability of self-injection locking laser of over 5.5 nm in wavelength near 1 μm. (c) Frequency tuning of the self-injection locking laser by sweeping PZT voltage.

Using delayed self-heterodyne interferometry (DSHI) [22], we can estimate the fundamental linewidth of the self-injection locking laser with a short delay line of 1 km long optical fiber. Of note, the actual linewidth and the stability of laser are further characterized in the following. The laser source signal is split into two paths by the coupler. One arm is offset with respect to the other of 200 MHz by an acousto-optic modulator (AOM, Chongqing Smart Science &Technology Development Co., Ltd, SGT200-02-N-1D). The other arm is delayed by 1 km single mode (SM) fiber. Finally, a coupler combines the two arms and fed into a photodetector (PD, EOT, ET3000). An electrical spectrum analyzer (ESA, Rohde &Schwarz, FSV 30) is used to analyze the beating signal acquired, as shown in Fig. 3 (a), where the signal light linewidth can be estimated from phase noise in the self-heterodyne RF beat signal [26]. In this case, the single-sided laser frequency noise is related to the power spectral density (PSD) of phase noise. Figure 3. (b) shows the PSD of the laser source. Consequently, the white frequency noise measures at a constant value of 13 Hz$^2$/Hz above 10 kHz of the frequency range, corresponding to a laser fundamental linewidth of 41 Hz.

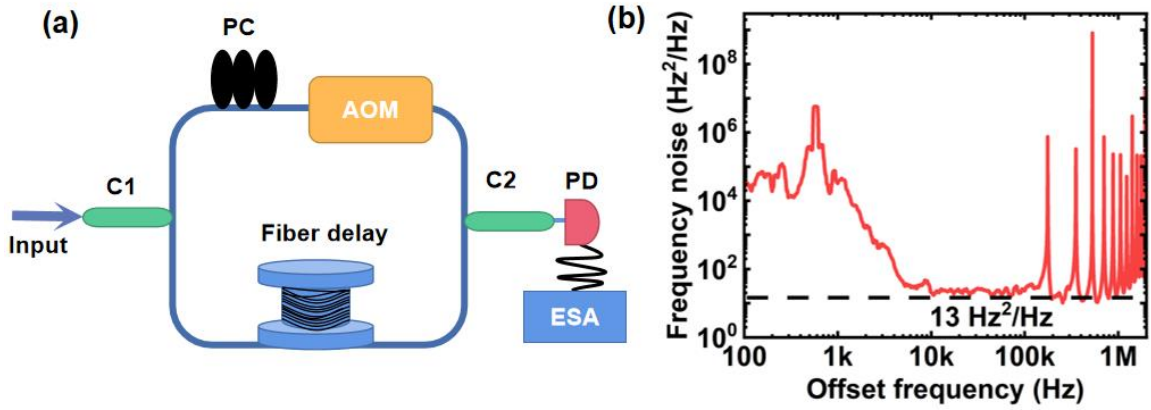

Fig. 3 Characterization of the self-injection locking laser's frequency stability. (a) Measurement setup. C1&C2, 50/50fiber coupler. AOM, acoustic optical modulator; ESA, electrical spectrum analyzer. (b) Characterization of the frequency noise measures at a constant value of 13 Hz$^2$/Hz above 30 kHz of the frequency range.

Finally, we characterize the laser frequency stability, the most common measurement technique is a spectral analysis of the laser through the beat-note measurement with a reference laser. The laser output is combined with the reference laser using a fiber coupler and attenuated to a suitable level before entering the photodetector and finally being recorded. The difference frequency between the two lasers is generated on the photodetector, the so-called beat frequency $f_{beat}$. For a precise measurement of the laser frequency stability, the reference laser should be much more stable than the laser under test. An ideal reference laser source would be a stabilized frequency comb [27-28], as this can provide both very high-frequency stability and a huge wavelength range. Figure 4.a. shows a schematic of the beat-note frequency generated by the self-injection locking laser and a stabilized frequency comb. The stabilized frequency comb is locked to a frequency-stabilized laser [29]. The linewidth of the optical comb is below 1 Hz, and the frequency stability is better than $10^{-15}$ at 1s [30]. The RF signal was analyzed with an ESA and a frequency counter. Figure 4.b. shows the linewidth of the beat-note frequency measurement was performed by an ESA with 300 Hz resolution bandwidth. Its Lorentz fitting characterizes a Full width at half maximum (FWHM) of 510.3 Hz. Fig. 4 c. shows that Allan deviation calculated from beat-note measurement between a self-injection locked laser and a stabilized frequency comb. The demonstrated level of frequency stability is good enough and Allan deviation of $10^{-11}$ or better at 1 s. In addition, we compared the Allen deviation of a commercial laser (NKT Y10 module) with ours. Our lasers show better performance on frequency stability compared to commercial lasers, which is mainly attributed to the high stability of the FP hollow resonator. At the same time there is still a lot of room for improvement in the FP hollow resonator, we expect that the noise performance of the laser can be improved by an order of magnitude via reduction of the intracavity optical loss, and improvement of the thermal stabilization of the resonator.

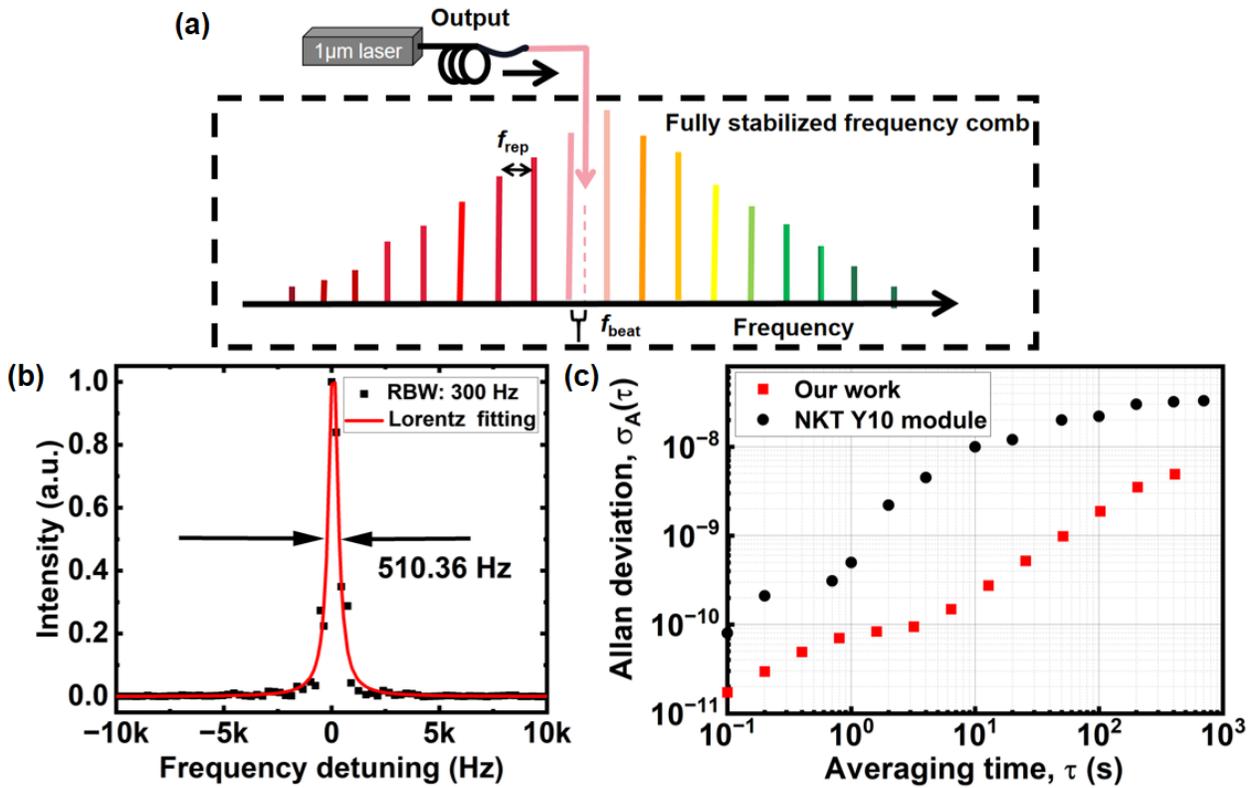

Fig. 4 The RF signal of the beat-note frequency is generated by beating a self-injection locking laser and a stabilized frequency comb. (a) Schematic of the beat-note frequency. (b) The linewidth of the beat-note frequency measurement was performed with 1 kHz resolution bandwidth. Points stand for the experimental data. The continuous red line is a Lorentzian fit of the data. (c) Allan deviation was calculated from the beat-note frequency.

In conclusion, we demonstrate a high-performance ultra-narrow linewidth laser source near 1 μm by self-injection locking of a laser chip to an FP hollow resonator of high-Q factor. A course tuning range over 5.5 nm is achieved by changing the driving current of the laser source. Meanwhile, a fine-tuning range of 373 MHz can be performed without mode hops, founding a good basis for further stabilization of laser frequencies and applications such as LIDAR. A 41 Hz ultra-narrow linewidth and a low-frequency noise level of 13 $Hz^2/Hz$ is obtained. For a detailed analysis of laser frequency stability, the laser output is combined with a stabilized frequency comb, and the result of the linewidth of the beat-note frequency is 510.3 Hz, with Allan deviation of $10^{-11}$ at 1 s. We also packaged a compact ultra-narrow linewidth 1μm laser that achieved a volume of less than 52.3ml. This laser module can be further reduced by improving the optical path arrangement and the performance of micro-optics. This work represents a substantial improvement in terms of quality factor, linewidth, and portability compared with ultra-narrow linewidth laser source near 1μm previously demonstrated with the high-Q resonator. This method can also be easily adapted to any other wavelength and fulfills the requirements of narrow-linewidth lasers in a compact and cost-effective way.

**ACKNOWLEDGMENTS**

This work was supported by Guangdong Major Project of Basic and Applied Basic Research (2020B0301030009), National Key R&D Program of China (2022YFA1205100, 2019YFA0705000, 2023YFB2805700), National Natural Science Foundation of China (62293523, 62075233, 12304421), Leading-edge technology Program of Jiangsu Natural Science Foundation (BK20192001), Zhangjiang Laboratory (ZJSP21A001), Program of Jiangsu Natural Science Foundation (BK20230770, BK20232033), CAS Project for Young Scientists in Basic Research (YSBR-69).

**AUTHOR DECLARATIONS**

Conflict of Interest The authors have no conflicts to disclose.

**DATA AVAILABILITY**

The data that support the findings of this study are available from the corresponding authors upon reasonable request.